\documentstyle[amssymb,aps,epsf,floats]{revtex}

\begin{document}
\title{Pairing fluctuations and pseudogaps in the \\
attractive Hubbard model}
\author{B. Kyung, S. Allen and A.-M. S. Tremblay$\dagger $}
\address{D\'{e}partement de physique and Centre de recherche sur\\
les propri\'{e}t\'{e}s \'{e}lectroniques de mat\'{e}riaux avanc\'{e}s.\\
$\dagger $Institut canadien de recherches avanc\'{e}es\\
Universit\'{e} de Sherbrooke, Sherbrooke, Qu\'{e}bec, Canada J1K 2R1}
\date{\today}
\maketitle

\begin{abstract}
The two-dimensional attractive Hubbard model is studied in the weak to
intermediate coupling regime by employing a non-perturbative approach. It is
first shown that this approach is in quantitative agreement with Monte Carlo
calculations for both single-particle and two-particle quantities. Both the
density of states and the single-particle spectral weight show a pseudogap
at the Fermi energy below some characteristic temperature $T^{\ast }$, also
in good agreement with quantum Monte Carlo calculations. The pseudogap is
caused by critical pairing fluctuations in the low-temperature renormalized
classical regime $\left( \hslash \omega <k_{B}T\right) $ of the
two-dimensional system. With increasing temperature the spectral weight
fills in the pseudogap instead of closing it and the pseudogap appears
earlier in the density of states than in the spectral function. Small
temperature changes around $T^{\ast }$ can modify the spectral weight over
frequency scales much larger than temperature. Several qualitative results
for the $s-$wave case should remain true for $d$-wave superconductors.

71.10.Fd, 71.27.+a, 71.10.-w,71.10.Pm.
\end{abstract}




\section{Introduction}

\label{section1}

For the past several years pseudogap phenomena found in the underdoped high
temperature superconductors\cite{StattTimusk:1999} and organic
superconductors\cite{Mayaffre:1994} have attracted considerable attention
among condensed matter physicists. For these materials the low frequency
spectral weight begins to be strongly suppressed below some characteristic
temperature $T^{\ast }$ that is higher than the transition temperature $%
T_{c} $. In the high temperature superconductors, this anomalous behavior
has been observed through various experimental probes such as photoemission 
\cite{Loeser:1996,Ding:1996}, specific heat\cite{Loram:1993}, tunneling\cite
{Renner:1998}, NMR\cite{Takigawa:1991}, and optical conductivity\cite
{Orenstein:1990}. Although various theoretical scenarios have been proposed,
there is no consensus at present. These proposals include spinon pair
formation without Bose-Einstein condensation of holons\cite
{Anderson:1987,Tanamoto:1992,Lee:1997}, stripes \cite
{Stripes:1989,Stripes:1994,Castellani:1995,Stripes:2000}, hidden $d-$density
wave order\cite{Laughlin:2000}, strong superconducting fluctuations \cite
{Doniach:1990,Emery:1995,Lotkev:2000-1997,Balseiro:1992,Beck:1991-1999,Franz:1998,Janko:1998,Kwon:1999,Varlamov:1999}%
, amplitude fluctuations with dimensional crossover\cite{Ranninger:2000} and
magnetic scenarios near the antiferromagnetic instability\cite
{Shen:1997,Schmalian:1998}.

Although the above theories for the origin of the pseudogap are very
different in detail, those that do not rest on spatial inhomogeneities can
be divided, roughly speaking, in two broad categories : Weak-coupling and
strong-coupling explanations. In the strong-coupling approaches, the
single-particle spectral weight is shifted to high-energies. There is no
weight at zero frequency at half-filling. That weight however increases as
one dopes away from half-filling, as qualitatively expected from the Physics
of a doped Mott insulator. Recent angle-resolved photoemission spectroscopy
(ARPES) experiments\cite{Ding:2000,Feng:2000} find, in the superconducting
state, a quasiparticle behavior consistent with this point of view. If we
consider instead a weak-coupling approach, either in the strict sense or as
an effective model for quasiparticles, the only known way of obtaining a
pseudogap is through coupling to renormalized classical fluctuations in two
dimensions. In the repulsive two-dimensional Hubbard model in the weak to
intermediate-coupling regime, analytical arguments\cite{Vilk:1996,Vilk:1997}
and detailed Monte Carlo simulations\cite{Moukouri:1999} strongly suggest
that indeed antiferromagnetic fluctuations can create a pseudogap in the
renormalized classical regime of fluctuations. This mechanism has been
confirmed recently by another approach\cite{Jarrell:1999} but earlier
studies had not found this effect.\cite{deisz:1996,white:1993}

The present paper focuses on superconducting fluctuations and the attractive
Hubbard model in weak to intermediate coupling. The purpose of the paper is
twofold. First, in section\ \ref{section2}, we validate, through comparisons
with Monte Carlo simulations, a non-perturbative many-body approach\cite
{Allen:2000} that is an extension of previous work on the repulsive model 
\cite{Vilk:1996,Vilk:1997}. Formal aspects of this method are presented in
the accompanying paper.\cite{Allen:2000} Then, in the second part of the
present paper (section\ \ref{section3}) we study the mechanism for pseudogap
formation due to superconducting fluctuations.\cite{Loktev:2000} More
extensive references on pseudogap formation in the attractive Hubbard model
may be found in section\ \ref{section3}. Earlier Monte Carlo work\cite
{VilkAllen:1998,Allen:2000} and analytical arguments \cite
{Vilk:1997,VilkAllen:1998} have suggested the appearance of a pseudogap in
the renormalized classical regime of pairing fluctuations. We study the
appearance of the pseudogap in both the density of states and the
single-particle spectral weight $A(\overrightarrow{k}_{F},\omega ),$ showing
that, in general, they occur at different temperatures. General comments on
the relation to pseudogap phenomena in high-temperature superconductors may
be found in the concluding paragraphs.

\section{A non-perturbative many-body approach compared with Monte-Carlo
results}

\label{section2}

In the first subsection, we present our approach in simple terms. More
formal arguments are in an accompanying paper\cite{Allen:2000}. In the
second subsection, we show that our approach is in quantitative agreement
with Monte Carlo simulations for both single-particle and two-particle
quantities.

\subsection{A non-perturbative sum-rule approach}

\label{section2a}

We consider the attractive Hubbard model for electrons on a two dimensional
square lattice 
\begin{equation}
H=\sum_{\vec{k},\sigma }\varepsilon _{\vec{k}}c_{\vec{k},\sigma }^{+}c_{\vec{%
k},\sigma }+\frac{U}{N}\sum_{\vec{k},\vec{p},\vec{q}}c_{\vec{k},\uparrow
}^{+}c_{\vec{k}+\vec{q},\uparrow }c_{\vec{p},\downarrow }^{+}c_{\vec{p}-\vec{%
q},\downarrow }\;,
\end{equation}
where $\varepsilon _{\vec{k}}=-2t(\cos k_{x}+\cos k_{y})$, $U$ is the
on-site attractive interaction $\left( U<0\right) $ and $N$ the number of
lattice sites. Throughout the calculations, the constants $t$, $\hbar $, $%
k_{B}$ and lattice spacing are taken to be unity. The index $\sigma $
represents spin. This Hamiltonian is not a valid model for $d$-wave
superconductors, but it is the simplest model for which it is possible to
check the accuracy of approximate many-body results against Monte Carlo
simulations. Once the accuracy of the many-body technique has been
established, it can be generalized to the $d-$wave case. Furthermore, many
qualitative results do not depend on whether one has $s-$wave or $d-$wave
pairing.

The non-perturbative approach to the attractive Hubbard model presented in
the accompanying paper is an extension of the approach used in the repulsive
case\cite{Vilk:1997}. In the first step (which was called zeroth order step
in the repulsive model case), the self-energy is obtained by a
Hartree-Fock-type factorization of the four-point function with the {\it %
additional constraint }that the factorization is exact when all space-time
coordinates coincide. It is important to note that this additional
constraint, analogous to the local field approximation of Singwi {\it et al.}
\cite{Singwi:1968,Vilk:1994}, leads to a degree of consistency between one-
and two-particle quantities that is absent from the standard Hartree-Fock
factorization. Functional differentiation, as in the Baym-Kadanoff approach 
\cite{Baym:1962}, then leads to a momentum- and frequency-independent
particle-particle irreducible vertex that satisfies \cite
{NoteAnsatzCanTransf} 
\begin{equation}
U_{pp}=U\frac{\langle (1-n_{\uparrow })n_{\downarrow }\rangle }{\langle
1-n_{\uparrow }\rangle \langle n_{\downarrow }\rangle }.  \label{Upp_ansatz}
\end{equation}
\begin{figure}%
%
\centerline{\epsfxsize 8cm \epsffile{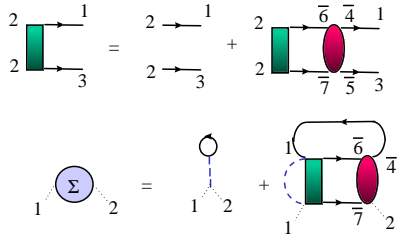}}%
%
\caption{The first line is a skeleton diagram representation of  the Bethe-Salpeter 
equation for the susceptibility in the particle-particle
channel  (Eq.(46) of accompanying paper\protect\cite{Allen:2000}) and the second line is the corresponding
equation for the self-energy (Eq.(58) of accompanying paper\protect\cite{Allen:2000}).
In the Hubbard model, the Fock contribution is absent, but in general it should be there.
Solid lines are Green's functions
and dashed lines represent the contact interaction U. The box and attached lines 
are the particle-particle susceptibility while the ellipse is the irreducible 
particle-particle vertex.
The particle-particle susceptibility is
obtained by identifying points $1$ and $3$ in the Bethe-Salpeter equation.}%
%
\label{fig0}%
%
\end{figure}%
%
With this approximation, the particle-particle susceptibility, which obeys
the Bethe-Salpeter equation illustrated on the first line of Fig.\ref{fig0},
can now be written as, 
\begin{equation}
\chi _{p}^{\left( 1\right) }(q)=\frac{\chi _{0}^{\left( 1\right) }(q)}{%
1+U_{pp}\chi _{0}^{\left( 1\right) }(q)}  \label{Chi_pp}
\end{equation}
where the irreducible particle-particle susceptibility is defined as 
\begin{equation}
\chi _{0}^{\left( 1\right) }(q)=\frac{T}{N}\sum_{k}G_{\sigma }^{\left(
1\right) }(q-k)G_{-\sigma }^{\left( 1\right) }(k)\;.  \label{eq14}
\end{equation}
The vertices and the Green functions in $\chi _{p}^{\left( 1\right) }(q)$
are at the same level of approximation in the sense that the irreducible
vertex $U_{pp}$ is obtained from the functional derivative of the
self-energy entering $G^{\left( 1\right) }$. The vertex $U_{pp}=\left(
\delta \Sigma /\delta G\right) $ is a constant and $\Sigma ^{\left( 1\right)
}$ in zero external field is also a constant, leading to a Green function
that has the same functional form as the non-interacting Green-function $%
G^{0}(k),$ where $k=\left( ik_{n},\vec{k}\right) $ stands for both the
fermionic Matsubara frequency $ik_{n}$ and the wave vector $\vec{k}.$ The
constant self-energy $\Sigma ^{\left( 1\right) }$ can be absorbed in the
chemical potential by working at constant filling. If needed, we have argued
in the accompanying paper\cite{Allen:2000} that the following should provide
a useful approximation for $\Sigma ^{\left( 1\right) }$ since it allows the
first-moment sum rule for the pair susceptibility to be satisfied \cite
{Allen:2000}, 
\begin{equation}
\Sigma ^{\left( 1\right) }\simeq \frac{U}{2}-\frac{U_{pp}\left( 1-n\right) }{%
2}  \label{Sigma(1)approx}
\end{equation}

At this first level of approximation, only the double occupancy ($D\equiv
\langle n_{\uparrow }n_{\downarrow }\rangle $) is needed to obtain the
irreducible vertex $U_{pp}$ and hence the pairing fluctuations. The value of 
$\langle n_{\uparrow }n_{\downarrow }\rangle $ can be borrowed from some
exact calculations or approximate estimates but, as in the repulsive case,
we found that accurate results are obtained when $\langle n_{\uparrow
}n_{\downarrow }\rangle $ is determined self-consistently from the following
``local-pair sum rule'', a consequence of the fluctuation-dissipation
theorem for the $s$-wave pairing susceptibility 
\begin{equation}
\frac{T}{N}\sum_{q}\chi _{p}(q)\exp (-iq_{n}0^{-})=\langle \Delta ^{\dagger
}\Delta \rangle =\langle n_{\uparrow }n_{\downarrow }\rangle ~.
\label{eq.SumRule}
\end{equation}
Substituting $\chi _{p}^{\left( 1\right) }$ Eq.(\ref{Chi_pp}) for the pair
susceptibility and Eq.(\ref{Upp_ansatz}) for the irreducible vertex $U_{pp}$
leads to an equation that determines double-occupancy, and hence $U_{pp},$
self-consistently 
\begin{equation}
\frac{T}{N}\sum_{q}\frac{\chi _{0}^{\left( 1\right) }(q)}{1+U\frac{\langle
(1-n_{\uparrow })n_{\downarrow }\rangle }{\langle 1-n_{\uparrow }\rangle
\langle n_{\downarrow }\rangle }\chi _{0}^{\left( 1\right) }(q)}\exp
(-iq_{n}0^{-})=\langle n_{\uparrow }n_{\downarrow }\rangle ~.
\label{DdeRegleSomme}
\end{equation}
This first part of the calculation is referred to as the Two-Particle
Self-Consistent (TPSC) approach.\cite{Note:TPSC}

Once the pair susceptibility has been found, as above, the next step of the
approach consists in improving the approximation for the single-particle
self-energy by starting from an exact expression where the high-frequency
Hartree-Fock behavior is singled out. This expression is represented by
skeleton diagrams on the second line of Fig.(\ref{fig0}). The piece of the
exact self-energy that is added to the Hartree-Fock behavior represents
low-frequency corrections. These involve Green functions, vertices and pair
susceptibility for which we already have good approximations from the first
step of the calculation. One thus substitutes in the exact expression the
irreducible low frequency vertex $U_{pp}$ as well as all the other
quantities at the same level of approximation, namely $G_{\sigma }^{\left(
1\right) }(k+q)$ and $\chi _{p}^{\left( 1\right) }(q)$ computed above,
obtaining in Fourier space 
\begin{equation}
\Sigma _{\sigma }^{\left( 2\right) }(k)=Un_{-\sigma }-U\frac{T}{N}%
\sum_{q}U_{pp}\chi _{p}^{\left( 1\right) }(q)G_{-\sigma }^{\left( 1\right)
}(q-k),  \label{Sigma(2)}
\end{equation}
where $q=\left( iq_{n},\overrightarrow{q}\right) $ stands for both the
bosonic Matsubara frequency and the wave vector. Here $T$ is the absolute
temperature. The resulting self-energy $\Sigma _{\sigma }^{\left( 2\right)
}(k)$ on the left hand-side is at the next level of approximation so it
differs from the self-energy entering the right-hand side. Physically, it is
a self-energy coming from taking into account cooperons. As stressed
previously\cite{Vilk:1997,Moukouri:1999}, it is important that the
irreducible vertex $U_{pp},$ (or $\Gamma ^{\left( 1\right) }$) as well as $%
G_{\sigma }^{\left( 1\right) }(k+q)$ and $\chi _{p}^{\left( 1\right) }(q)$
all be at the same level of approximation otherwise some results, in
particular with regards to the pseudogap, may come out qualitatively wrong.

The particle-particle irreducible vertex $U_{pp}$ may be regarded as the
renormalized interaction strength containing vertex corrections. Note that
in the expression for the self-energy $\Sigma ^{\left( 2\right) }$, Eq.(\ref
{Sigma(2)}), one of the vertices is bare while the other one is dressed with
the same\cite{Note1} particle-particle irreducible vertex function $U_{pp}$
that appears in the paring susceptibility. In other words, we do not assume
that a Migdal theorem applies. If this had been the case, the self-energy
would have, amongst other things, been proportional to $U^{2}$ instead of $%
UU_{pp}.$ In $T$-matrix theory the bare $U$ is used everywhere which, in
particular, leads to a finite-temperature phase transition at the mean-field
temperature. In our case, as described below, we proceed differently,
avoiding altogether a finite-temperature phase transition in two-dimensions.

We briefly summarize some of the constraints satisfied by the above
non-perturbative approach\cite{Allen:2000}. In Eq.(\ref{eq.SumRule}), $%
\Delta $ is the local $s-$wave order parameter $c_{i\downarrow }c_{i\uparrow
}$. Anti-commutation relations, or equivalently the Pauli principle, imply
that $\langle \left[ \Delta ,\Delta ^{\dagger }\right] \rangle =1-n.$ This
in turn means that the convergence factor $\exp (-iq_{n}0^{-})$ in the
local-pair sum rule is necessary because $\langle \left[ \Delta ,\Delta
^{\dagger }\right] \rangle =1-n$ implies that, except at $n=1$, one needs to
specify if $\tau =0^{+}$ or $\tau =0^{-}$ in the imaginary-time pair
susceptibility. Either one of these limits however leads to the same value
of $U_{pp}$ since our approach\cite{Allen:2000} satisfies exactly this
consequence of the Pauli principle\cite{Note:Moment0}: $\langle \left[
\Delta ,\Delta ^{\dagger }\right] \rangle =1-n$. In complete analogy with
the repulsive case discussed in Ref.\cite{Vilk:1994}, one can invoke the 2D
phase space factor, $2\pi qdq$, and the Ornstein-Zernicke form of the
pairing correlation function near a critical point to show the following.
Deep in the renormalized classical regime, where the characteristic
frequency $\nu _{c}$ of the retarded pairing susceptibility satisfies $\nu
_{c}\lesssim T,$ the superconducting correlation length $\xi $ increases
exponentially $\xi \sim exp(C/T)$ with decreasing temperature. In the latter
expression, $C$ may be temperature dependent. This behavior is the one
expected in the $O\left( n=\infty \right) $ universality class \cite
{Dare:1996} and not in the $XY$, or $O\left( 2\right) ,$ universality class,
where $\xi _{BKT}\sim exp(C/\left( T-T_{BKT}\right) ^{1/2}).$ The precise
dependence on temperature of the correlation length in the temperature range
between the beginning of the renormalized-classical regime and the actual
critical regime is not known analytically. In the regime that we explore,
the pseudogap begins to open at a temperature $T$ that is quite a bit larger
than $T_{BKT}.$ The latter was estimated to be at most of order of $0.1t$
for $|U|=4$ by Moreo {\it et al.} \cite{Moreo:1991}. As in the repulsive
case, we will see below that the pseudogap appears in $A(\overrightarrow{k}%
_{F},\omega )$ if the pairing correlation length grows faster with
decreasing temperature than the single-particle thermal de Broglie
wavelength $\xi _{th}=v_{F}/T.$

\subsection{Comparisons with Monte Carlo calculations}

In this section we show, by comparing with Monte Carlo calculations, that
the present non-perturbative approach is an accurate approximation. The
calculations are performed for the same lattice size as the corresponding
Monte Carlo calculations. It is important to note that at half-filling the
Lieb-Mattis canonical transformation $c_{i\downarrow }\rightarrow \exp
\left( -i\overrightarrow{Q}{\bf \cdot }\overrightarrow{r}_{i}\right)
c_{i\downarrow }^{\dagger }$, with $\overrightarrow{Q}=\left( \pi ,\pi
\right) $, maps the attractive model onto the repulsive one, pair
fluctuations at wave vector $\overrightarrow{q}$ being mapped onto
transverse spin fluctuations at wave vector $\overrightarrow{q}+%
\overrightarrow{Q}.$ With the {\it proviso} that in the attractive model at
half-filling one would need, because of symmetry\cite{Moukouri:1999}\cite
{Allen:1999}, to take into account the charge fluctuations, we can state
that the comparisons done in the repulsive case\cite{Vilk:1997}\cite
{Moukouri:1999} apply for the canonically equivalent attractive case. We
note in particular that it was shown that the convergence to the
infinite-size limit is similar in the Monte Carlo and in the
non-perturbative approach\cite{Moukouri:1999}. We restrict our discussion to
cases away from half-filling. The Quantum Monte Carlo simulations\cite
{hirsh:1985}\cite{white:1989} that we performed were done using a Trotter
decomposition with increment $\Delta \tau =1/10$ in imaginary time and the
determinantal approach\cite{Blankenbecler:1981}. Typically, about $10^{5}$
or more Monte Carlo sweeps of the space-time lattice are performed.

We begin with double occupancy. Table I shows the value of double occupancy
calculated from Monte Carlo simulations for various temperatures and
fillings. The last column shows the value obtained from the self-consistent
equation for double-occupancy Eq.(\ref{DdeRegleSomme}).%
\begin{figure}%
%
\centerline{\epsfxsize 8cm \epsffile{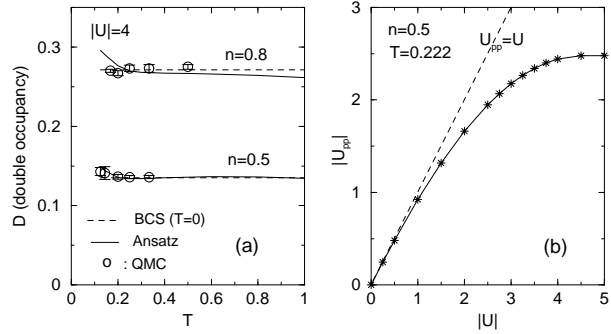}}%
%
\caption{(a) Double occupancy D calculated from the ansatz (solid curves),
from the BCS ground state (dashed lines), and from the QMC simulations
(circles) for $|U|=4$ at $n=0.5$ and 0.8. (b) Renormalized interaction
strength $|U_{pp}|$ as a function of bare $|U|$ for $T=0.222$ at $n=0.5$.
The dashed line is $U_{pp}=U$.}%
%
\label{fig1}%
%
\end{figure}%
%
\ 
\begin{table}%
%
\caption{Double occupancy for various fillings $n$ and temperature $\beta$ obtained
from QMC calculations and from the TPSC approach. The systematic error in the QMC
calculation is of order $1\%$, larger than the statistical uncertainty}
%

\begin{tabular}{lllcc}
$U$ & $n$ & $\beta $ & QMC & TPSC \\ 
$-4$ & $0.8$ & $4$ & \multicolumn{1}{l}{$0.2671\pm 10^{-4}$} & 
\multicolumn{1}{l}{$0.273$} \\ 
$-4$ & $0.8$ & $6$ & \multicolumn{1}{l}{$0.2703\pm 10^{-4}$} & 
\multicolumn{1}{l}{$0.288$} \\ 
$-4$ & $0.55$ & $4$ & \multicolumn{1}{l}{$0.1561\pm 10^{-4}$} & 
\multicolumn{1}{l}{$0.157$} \\ 
$-4$ & $0.55$ & $6$ & \multicolumn{1}{l}{$0.1590\pm 10^{-4}$} & 
\multicolumn{1}{l}{$0.169$}
\end{tabular}
\end{table}%
%
Clearly the accuracy of the approximation deteriorates at low temperature.
This is illustrated by Fig.\ref{fig1}(a) where, for both densities studied,
the solid line starts to deviate from the Monte Carlo data around $\beta <5.$
As in the repulsive case, this occurs because the self-consistent expression
for double-occupancy Eq.(\ref{DdeRegleSomme}) fails once we enter the
renormalized-classical regime where a pseudogap appears. Following Ref.\cite
{Vilk5.6:1997}, we expect that the pseudogap in $A(\overrightarrow{k}%
_{F},\omega )$ opens up as a precursor of Bogoliubov quasiparticles in the $%
BSC$ ground state. Since this ground state starts to control the Physics, it
is natural to expect that a high-energy quantity such as $D$ should, in the
pseudogap region, take the zero temperature $BCS$ value, namely 
\begin{equation}
\langle n_{\uparrow }n_{\downarrow }\rangle =(\frac{1}{N}\sum_{\vec{k}}v_{%
\vec{k}}^{2})^{2}+(\frac{1}{N}\sum_{\vec{k}}u_{\vec{k}}v_{\vec{k}})^{2}\;,
\end{equation}
where 
\begin{eqnarray}
u_{\vec{k}}^{2} &=&\frac{1}{2}(1+\frac{\varepsilon _{\vec{k}}-\mu }{E_{\vec{k%
}}})  \nonumber \\
v_{\vec{k}}^{2} &=&\frac{1}{2}(1-\frac{\varepsilon _{\vec{k}}-\mu }{E_{\vec{k%
}}})  \nonumber \\
E_{\vec{k}} &=&\sqrt{(\varepsilon _{\vec{k}}-\mu )^{2}+\Delta ^{2}}\;.
\label{eq17}
\end{eqnarray}
with $\Delta $ the BCS mean-field gap. The chemical potential $\mu $ and gap 
$\Delta $ are determined self-consistently through the number and gap
equations for given $U$, $T$ and $n$. The value of $D$ obtained with this
approach is plotted as a dotted line in Fig.\ref{fig1}(a), where it is
apparent that the agreement with Monte Carlo calculations is excellent. In
fact, for $\left| U\right| >3,$ the agreement is always at the few percent
level. For smaller $\left| U\right| $, deviations occur, probably because at
small coupling the order parameter at $\vec{q}{\bf =}0$ does not dominate
anymore the sum over all wave vectors in $1/N\sum_{\vec{q}}\langle \Delta (%
\vec{q})^{\dagger }\Delta (-\vec{q})\rangle =\langle \Delta ^{\dagger
}\Delta \rangle =\langle n_{\uparrow }n_{\downarrow }\rangle $. In fact, as
we shall see below, a good estimate of double-occupancy may also be obtained
at small $U$ just from second-order perturbation theory. Now consider the
temperature dependence of double-occupancy. The dependence predicted by the
finite-temperature BCS result is on a scale $T=1$ at $|U|=4.$ That
temperature dependence is clearly wrong for our problem since above $T_{MF}$
the BCS approach would give us back the non-interacting value. Hence the BCS
result may be used only in the following way. For $\left| U\right| >3$ we
can use the $T=0$ value of $\langle n_{\uparrow }n_{\downarrow }\rangle $ in
the pseudogap region and the self-consistent value{\it \ }Eq.(\ref
{Upp_ansatz}) above it. In the general case, $\langle n_{\uparrow
}n_{\downarrow }\rangle $ does depend on temperature in the pseudogap
regime, but that dependence should be relatively weak, as discussed in the
repulsive case.\cite{Vilk:1997}

We stress, however, that deep in the pseudogap regime our approach becomes
eventually inaccurate when we obtain $\langle n_{\uparrow }n_{\downarrow
}\rangle $ from the self-consistent equation (\ref{DdeRegleSomme}). The
reason for the loss of accuracy is analogous to that found at $n=1$ in the
repulsive\cite{Vilk:1997} case: In the present $U<0$ case, $\langle
n_{\uparrow }n_{\downarrow }\rangle \rightarrow \langle n_{\downarrow
}\rangle $ as $T\rightarrow 0$ to prevent a finite temperature phase
transition. The approach also eventually becomes less accurate in the
pseudogap regime when we take $\langle n_{\uparrow }n_{\downarrow }\rangle $
from BCS, but the fact that a more physically reasonable value of $\langle
n_{\uparrow }n_{\downarrow }\rangle $ may be obtained in that case at $T=0$
helps extrapolate a little bit deeper in the pseudogap regime. The internal
accuracy check discussed at the end of this section helps quantify the
region of validity of the approach.

Before moving on with the comparisons, a few comments on the actual
renormalized interaction strength $|U_{pp}|$ resulting from the Two-Particle
Self-Consistent calculation. In Fig.\ref{fig1}(b) $|U_{pp}|$ (denoted as
stars) is plotted for $T=0.222$ as a function of bare $U$ by using the
self-consistent expressions Eqs.(\ref{Upp_ansatz}) and (\ref{DdeRegleSomme}%
). For this temperature, $U_{pp}$ approaches bare $U$ for $|U|\leq 1$ while
in the intermediate coupling regime $1\leq |U|<\mbox{bandwidth}$ one notices
a strong deviation from the bare $\left| U\right| .$ This deviation of $%
U_{pp}$ from $U$ makes a drastic difference, in particular in the
two-particle function $\chi _{p}\left( \vec{q}\right) $ which ultimately
governs the particle dynamics, including the pseudogap behavior. The
saturation of $\left| U_{pp}\right| $ in Fig.\ref{fig1}(b) signals the onset
of the strong-coupling regime inaccessible in our approach that does not
take into account the strong frequency dependence of vertices and
self-energy present in this limit.

\begin{figure}%
%
\centerline{\epsfxsize 8cm \epsffile{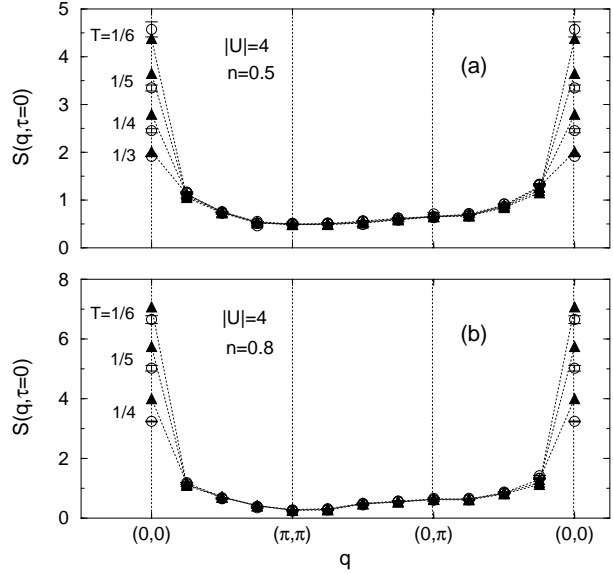}}%
%
\caption{Calculated s-wave paring structure factor $S(\vec{q},\protect\tau
=0)$ (filled triangles) and QMC $S(\vec{q},\protect\tau =0)$ (circles) for $|U|=4$ 
and various temperatures.  
(a) at $n=0.5$ and (b) at $n=0.8$ on a $8\times 8$ lattice. The 
dashed lines are to guide the eye..}%
%
\label{fig3}%
%
\end{figure}%
%
Our approach gives a very accurate result for double-occupancy, but what
about correlation functions? Another two-particle quantity related to the
susceptibility is the pair structure factor. In Fig.\ref{fig3} the
calculated s-wave pair structure factor (filled triangles) $S(\vec{q},\tau
=0)\equiv $ $\langle \Delta (\vec{q})\Delta ^{\dagger }(-\vec{q})+\Delta (%
\vec{q})^{\dagger }\Delta (-\vec{q})\rangle ,$ where $\Delta (\vec{q})^{+}=%
\frac{1}{\sqrt{N}}\sum_{\vec{k}}c_{\vec{q}-\vec{k},\uparrow }^{+}c_{\vec{k}%
,\downarrow }^{+},$ is compared with our QMC results (circles) for $|U|=4$
at $n=0.5$ (top panel) and $0.8$ (bottom panel) on a $8\times 8$ lattice.\
As the temperature decreases, the $\vec{q}=0$ mode becomes more singular in
both results, a characteristic feature for growing s-wave pairing
fluctuations. In most of the Brillouin zone, the agreement is excellent, in
particular, for $n=0.5$ where the maximum difference is less than $10\%$
(Fig.~\ref{fig3}(a)). For $n=0.8$ our calculated structure factor
overestimates QMC results at most by $20\%$.

\begin{figure}%
%
\centerline{\epsfxsize 8cm \epsffile{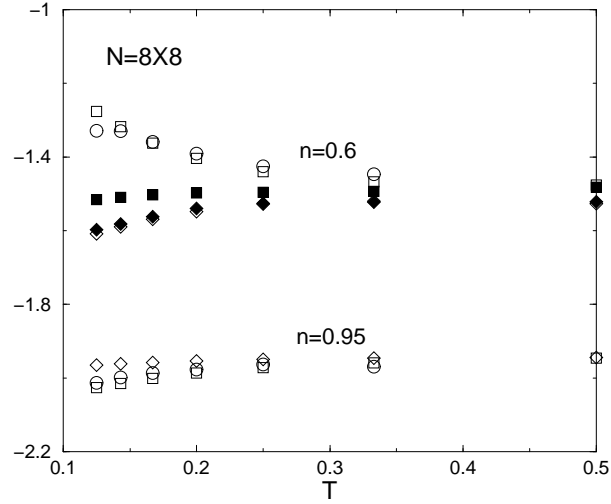}}%
%
\caption{Comparison of the chemical potential shifts $\mu^{(1)}-\mu_{0}$ 
(open diamonds) and $\mu^{(2)}-\mu_{0}$ (open squares) with the results
of QMC calculations (open circles), all done on a $8 \times 8$ lattice. QMC error is 
a few percent, smaller than the open circles.
Upper set of points are for $n=0.6$, and lower set of points are for $n=0.95$. 
For $n=0.6$, many-body calculations for a $64 \times 64$ lattice are also illustrated by
filled squares for $\mu^{(2)}-\mu_{0}$ , and by filled diamonds for $\mu^{(1)}-\mu_{0}$.
In all cases, $\mu^{(2)}-\mu_{0}$ is closer to QMC data than $\mu^{(1)}-\mu_{0}$.}%
%
\label{fig3x}%
%
\end{figure}%
%
At the first level of approximation, we can also estimate the
interaction-induced shift in chemical potential by starting from our
approximate expression for the self-energy $\Sigma ^{\left( 1\right) }$, Eq.(%
\ref{Sigma(1)approx}). Let us call the corresponding chemical potential $\mu
^{\left( 1\right) }=\mu _{0}+\Sigma ^{\left( 1\right) }.$ Our best estimate
of single-particle quantities is obtained from the self-energy at the second
level of approximation, Eq.(\ref{Sigma(2)}). The corresponding chemical
potential $\mu ^{\left( 2\right) }$ is calculated by requiring that the
filling be the same as the one used at the first level of approximation.
This procedure is identical to that for the repulsive model\cite
{Vilk(47):1997} and was suggested by Luttinger. It is discussed also in
Sec.IV-C of the accompanying paper.\cite{Allen:2000} Figure~\ref{fig3x}
illustrates how the two estimates for the chemical potential on a $8\times 8$
lattice converge towards the value obtained from QMC calculations for the
same size lattice. Consider the data for filling $n=0.6,$ in the upper part
of the figure. The open squares, representing $\mu ^{\left( 2\right) }-\mu
_{0},$ agree, within the error bars, with the QMC data represented by open
circles. The first estimate for the chemical potential shift $\mu ^{\left(
1\right) }-\mu _{0}$ (open diamonds) starts to deviate from both the QMC
data (open circles) and from $\mu ^{\left( 2\right) }-\mu _{0}$ (open
squares) at the temperature where the pseudogap opens up (see Fig.\ref{fig8}
below). Below this temperature, the self-energy becomes strongly frequency
and momentum dependent, a feature captured by our improved estimate $\Sigma
^{\left( 2\right) }$ for the self-energy, but not by our first estimate, $%
\Sigma ^{\left( 1\right) }.$ The filled squares ($\mu ^{\left( 2\right)
}-\mu _{0}$) and filled diamonds ($\mu ^{\left( 1\right) }-\mu _{0}$) were
obtained for a $64\times 64$ lattice. They illustrate that one should
compare finite-size QMC calculations to many-body calculations done on same
size systems. They also illustrate that the second estimate for the chemical
potential shift (squares) is more sensitive to system size. This is expected
from the fact that it is only at the second level of approximation that the
pseudogap appears for large correlation lengths. Let us now move to the
lower part of Fig.\ref{fig3x}, where results closer to half-filling are
plotted. The chemical potential there is very close to its exact
temperature-independent half-filling value, $U/2,$ hence there is not much
room to see the difference between the first and second estimate for the
chemical potential. Nevertheless, even for $n=0.95$, the second estimate
(open squares) is closer to the QMC data (open circles) than the first
estimate (open diamonds).

We now compare the momentum distribution, a static quantity, obtained from
QMC and from our analytical approach at the second level of approximation $%
\Sigma ^{\left( 2\right) }$. 
\begin{figure}%
%
\centerline{\epsfxsize 8cm \epsffile{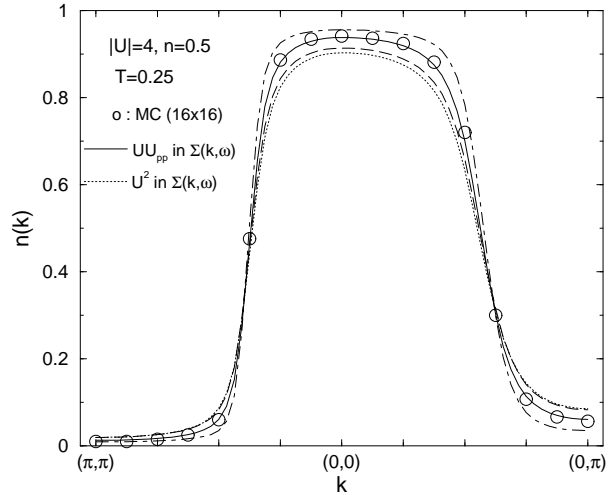}}%
%
\caption{The momentum dependent occupation number $n(\vec{k})$ for $|U|=4$
and $n=0.5$ at $T=0.25$. The
circles denote $n(\vec{k})$ from QMC calculations by Trivedi {\it et al.}
\protect\cite{Trivedi:1995} on a $16\times16$ lattice. 
The solid curve is calculated according to the
equations given in this paper, while the dashed one is computed by replacing
$U_{pp}$ by $U$ in the self-energy with all the rest unchanged. 
The long-dash
line is the result of a self-consistent T-matrix calculation, and the dot-dash line
the result of second-order perturbation theory.}%
%
\label{fig4}%
%
\end{figure}%
%
\ The momentum dependent occupation number $n(\vec{k})$ (solid curve) is
plotted in Fig.~\ref{fig4} along with the QMC calculations (circles) by
Trivedi {\it et al.}\cite{Trivedi:1995} for a $16\times 16$ lattice, in a
regime where the size dependence is negligible. The momentum distribution $n(%
\vec{k})$ drops rapidly near the Fermi surface, corroborating that for $%
|U|=4 $ the electrons are in the degenerate state, instead of in the
non-degenerate state of the strong coupling regime or of the preformed-pair
scenario. The agreement between the non-perturbative method and QMC is
clearly excellent. If we had assumed a Migdal theorem and taken $U^{2}$
instead of $UU_{pp}$ in Eq.(\ref{Sigma(2)}) for the self-energy, then we
would have obtained the dotted curve, which in absolute value differs as
much from the QMC result as a non-interacting Fermi distribution would.
Clearly Migdal's theorem does not apply for this problem. In addition, Fig.%
\ref{fig4} also shows (long dashes) that a self-consistent $T-$matrix
calculation does not compare to Monte Carlo data as well as our approach.
Similarly, second-order perturbation theory (dot-dash) does not do well.

\begin{figure}%
%
\centerline{\epsfxsize 8cm \epsffile{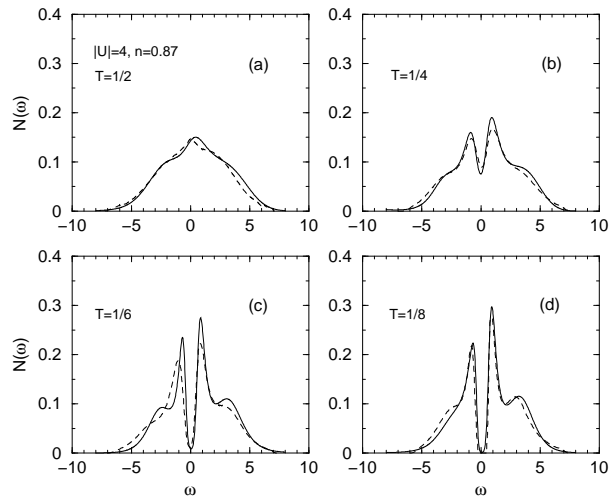}}%
%
\caption{The single-particle density of states for $|U|=4$ and $n=0.87$ at
different temperatures on a $8\times 8$ lattice. (a) $T=1/2$, (b) $T=1/4$,
(c) $T=1/6$, and (d) $T=1/8$. Both the solid line obtained from many-body calculations
and the dashed line taken from the QMC calculations of 
Ref.\protect\cite{Moreo:1992} are obtained by analytic continuation 
of imaginary-time data using maximum entropy. The absolute error 
chosen for the Maximum Entropy continuation of the imaginary time 
many-body Green function for panels (a) to (d) is, respectively, $0.003,
0.004, 0.003, 0.001$.}%
%
\label{fig2}%
%
\end{figure}

In Fig.~\ref{fig2} we present the total density of states $N^{-1}\sum_{%
\overrightarrow{k}}A(\overrightarrow{k},\omega )$ for $U=4$ and $n=0.87$ on
an $8\times 8$ lattice and compare with existing QMC calculations by Moreo 
{\it et al.}\cite{Moreo:1992}%
%
\ The many-body non-perturbative calculation is done in Matsubara frequency
and analytically continued to real frequency using the same Maximum Entropy
(ME) technique that is used for QMC calculations.\cite{ME:1996} This allows
us to smooth out the results in the same way as in the QMC calculation,
namely by including statistical uncertainties in our imaginary-time data.
This procedure was discussed in Ref.\cite{Moukouri:1999}. The uncertainties
in the QMC data\cite{Moreo:1992} that are presented in Fig.\ref{fig2} were
not quoted. Hence, we chose the statistical uncertainties in the
corresponding many-body calculations in such a way that the calculations
presented in Fig.\ref{fig2}(d) have the same degree of smoothness as the
corresponding QMC data. More specifically, these uncertainties are of order $%
0.003$ on the absolute value of the imaginary time Green functions, which is
typical of Monte Carlo calculations. At $T=1/2$ the density of states is
similar to that for the non-interacting system. At $T\leq 1/4$, however, the
spectral weight near the Fermi energy begins to be suppressed significantly
with decreasing temperature, leading to a pseudogap. The small shoulders in
the intermediate frequency regime for $T=1/6$ and $1/8$ come from finite
size effect. When we use a $64\times 64$ lattice, these shoulders completely
disappear.

In the case of the single-particle spectral weight, the latest QMC
calculations at $n=1$\cite{Moukouri:1999} included studies of the
finite-size effects, of the imaginary-time discretisation and of the
uncertainties induced by the size of the Monte Carlo sample. They have shown
that, at half-filling, there is indeed a pseudogap, in contrast to earlier
findings \cite{white:1993}. Detailed comparisons with the many-body approach
analog to the present one have been done. Although these studies were for
the repulsive model, at half-filling the results apply for the attractive
model since they are canonically equivalent at $n=1.$ One only needs to
generalize the many-body approach to include the presence of the $SO\left(
3\right) $ symmetry, as done in Ref.\cite{Moukouri:1999}. Slightly away from
half-filling, namely for $n=0.95,$ QMC simulations\cite{Allen:1999}\cite
{Allen:2000b} have found that a similar pseudogap opens up for $U=-4$ at a
temperature large enough that the $SO\left( 3\right) $ symmetry present at
half-filling is barely broken $\left( T_{X}\gg \mu \right) .$

\begin{figure}%
%
\centerline{\epsfxsize 8cm \epsffile{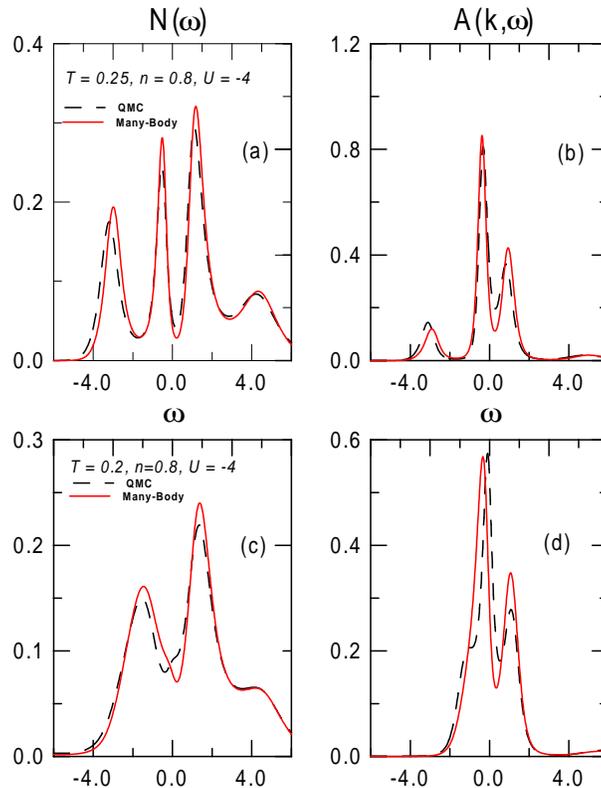}}%
%
\caption{Total density of states (first column) and single-particle
spectral weight at wave vector $(0,3\pi/4)$ (second column) 
are shown for $U=-4, n=0.8$ and
two different temperatures. The lattice size is $8 \times 8$. Dashed lines are
the result of Maximum Entropy continuation of QMC data, and solid lines the result
of Maximum Entropy continuation of many-body results with the same errors added in.  
Plots (a) and (b) on the first line are, respectively, $N(\omega)$ the total density
of states, and $A(0,3\pi/4,\omega)$  the single-particle spectral weight at a point
near the Fermi surface, for $T=0.25$. 
Panels (c) and (d) on the last line are the same plots but for $T=0.2$. 
The number of measurements done is $1.2 \times 10^5 $ for $T=0.25$
and  $1.6 \times 10^5 $for $T=0.2$. In all cases, the absolute statistical
error on the imaginary-time data is of order $2.0 \times 10^{-3}$. Singular values less than 
$10^{-3}$ are dropped in the Maximum Entropy inversion.}%
%
\label{fig_derniere}%
%
\end{figure}

\begin{figure}%
%
\centerline{\epsfxsize 8cm \epsffile{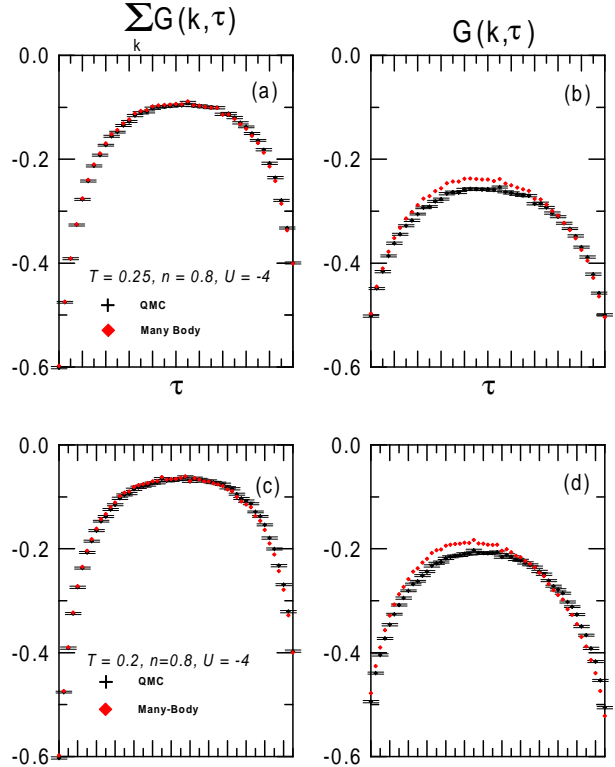}}%
%
\caption{Imaginary-time data from which are extracted the density of states
and spectral weights appearing in the respective panels of Fig.\ref{fig_derniere}.
Error bars are shown only for the Monte Carlo data but the analytical data has
been added the same errors as the Monte Carlo data.}%
%
\label{fig_derniere_bis}%
%
\end{figure}

Here we present results for $A(\overrightarrow{k},\omega )$ and for the
total density of states $N\left( \omega \right) =N^{-1}\sum_{\overrightarrow{%
k}}A(\overrightarrow{k},\omega )$ at $n=0.8,$ $U=-4$ and two different
temperatures, $T=0.25$ and $T=0.2,$ in the pseudogap regime. For that
filling, finite-size studies are complicated by the fact that the available
wave vectors do not necessarily coincide with the non-interacting Fermi
surface. The closest such wave vector, $\overrightarrow{k}=\left( 0,3\pi
/4\right) $, is the one chosen for comparisons of $A(\overrightarrow{k}%
,\omega )$. The results of QMC calculations are shown in Fig.(\ref
{fig_derniere}) as dashed lines and those of the many-body approach as solid
lines for the same $8\times 8$ system size. 
%
\ The imaginary-time data corresponding to Fig.(\ref{fig_derniere}) appears
in Fig.(\ref{fig_derniere_bis}).%
%
$\ $It is clear that all the main features of the Monte Carlo are seen by
the many-body approach. The relative size of the split peaks at the Fermi
surface is very sensitive to the actual location of the Fermi wave vector.

We conclude this section with the accuracy check described in the
accompanying paper\cite{Allen:2000} and in previous work\cite{Vilk:1997}.
The question here is whether it is possible to find the domain of validity
of the approach even in the absence of Monte Carlo data. The answer is given
by Table II, where all results were computed for a $64\times 64$ lattice.
One should focus on the three columns labeled {\it Many-Body. }In the first
of these, one finds the value of $U\left\langle n_{\uparrow }n_{\downarrow
}\right\rangle $ computed with TPSC, {\it i.e.} from{\it \ }Eq.(\ref
{Upp_ansatz}) and the local-pair sum rule Eq.(\ref{DdeRegleSomme}). That
number is the same as that which would be obtained from {\rm Tr}$\left[
\Sigma ^{\left( 2\right) }G^{\left( 1\right) }\right] .$ However, finding by
how much it differs from {\rm Tr}$\left[ \Sigma ^{\left( 2\right) }G^{\left(
2\right) }\right] $, listed in the second column labeled {\it Many-Body},
gives an indication of how much the theory is internally consistent. The
third column labeled {\it Many-Body }gives the absolute value of the
relative difference between the first two {\it Many-Body }columns and it
helps appreciate where the theory fails. Clearly, as temperature decreases
and one enters the renormalized classical regime, the theory becomes invalid
since the difference between $U\left\langle n_{\uparrow }n_{\downarrow
}\right\rangle $ and {\rm Tr}$\left[ \Sigma ^{\left( 2\right) }G^{\left(
2\right) }\right] $ starts to increase rapidly. As expected also, the theory
is better at smaller coupling. Instead of computing $U\left\langle
n_{\uparrow }n_{\downarrow }\right\rangle $ with the TPSC approach Eq.(\ref
{DdeRegleSomme}), one may also take it from the zero-temperature BCS value
as described above. Comparing with the corresponding {\it Many-Body }%
columns, it is clear the BCS estimate of double occupancy does not compare
as well with the TPSC result at small coupling, as explained above. In fact,
in this region, second-order perturbation theory compares better. This is
partly because the Hartree-Fock contribution to double-occupancy is
dominant. The worse results for perturbation theory are for quarter filling
at $U=-4$. Near half-filling it is known that second-order perturbation
theory works well again. Note also that even if we start from the BCS result
for $U\left\langle n_{\uparrow }n_{\downarrow }\right\rangle $, one can tell
that the theory is becoming less accurate at too low temperature because,
there again, $U\left\langle n_{\uparrow }n_{\downarrow }\right\rangle $ and 
{\rm Tr}$\left[ \Sigma ^{\left( 2\right) }G^{\left( 2\right) }\right] $
begin to differ more and more from each other. This internal accuracy check,
however, cannot tell us which approach is more accurate compared with QMC.%
\begin{table}%
%
\caption{Internal consistency check, for various $U$, $n$, and $T$ on a
$64 \times 64$ lattice. The three columns 
labeled {\it Many-Body} give, respectively, the 
value of $U<n_{\uparrow}n_{\downarrow}>$ obtained at the first
level of approximation with the TPSC
approach,  
that obtained at the second level of approximation from 
$Tr[\Sigma^{(2)}G^{(2)}]$ and the absolute value of the difference between
the two results. The following column is {\rm Tr}$\left[ \Sigma G\right] $
obtained from second-order perturbation theory, and the last three columns
are analogous to those where double-occupancy is obtained
from the TPSC but they start from the $T=0$ BCS estimate of
$U<n_{\uparrow}n_{\downarrow}>$.}
%

\begin{tabular}{cccccccccc}
&  &  & {\it Many-Body} & {\it Many-Body} & {\it Many-Body} & $2^{nd}order$
& BCS & BCS & BCS \\ 
$U$ & $n$ & $T$ & $U\left\langle n_{\uparrow }n_{\downarrow }\right\rangle $
& {\rm Tr}$\left[ \Sigma ^{\left( 2\right) }G^{\left( 2\right) }\right] $ & $%
\left| diff\%\right| $ & {\rm Tr}$\left[ \Sigma G\right] $ & $U\left\langle
n_{\uparrow }n_{\downarrow }\right\rangle $ & {\rm Tr}$\left[ \Sigma
^{\left( 2\right) }G^{\left( 2\right) }\right] $ & $\left| diff\%\right| $
\\ 
$-2$ & $0.5$ & $0.1$ & $-0.1887$ & $-0.1838$ & $2.63$ & $-0.1728$ & $-0.1376$
& $-0.1366$ & $0.74$ \\ 
$-2$ & $0.5$ & $0.2$ & $-0.1892$ & $-0.1861$ & $1.61$ & $-0.1743$ & $-0.1376$
& $-0.1371$ & $0.39$ \\ 
$-2$ & $0.5$ & $0.3$ & $-0.1913$ & $-0.1888$ & $1.29$ & $-0.1762$ & $-0.1376$
& $-0.1372$ & $0.26$ \\ 
$-2$ & $0.5$ & $0.4$ & $-0.1939$ & $-0.1917$ & $1.15$ & $-0.1787$ & $-0.1376$
& $-0.1373$ & $0.19$ \\ 
$-2$ & $0.5$ & $0.5$ & $-0.1963$ & $-0.1942$ & $1.08$ & $-0.1812$ & $-0.1376$
& $-0.1374$ & $0.16$ \\ 
$-2$ & $0.5$ & $1.0$ & $-0.1988$ & $-0.1970$ & $0.88$ & $-0.1875$ & $-0.1376$
& $-0.1375$ & $0.10$ \\ 
$-2$ & $0.8$ & $0.1$ & $-0.4262$ & $-0.4196$ & $1.56$ & $-0.4045$ & $-0.3722$
& $-0.3696$ & $0.67$ \\ 
$-2$ & $0.8$ & $0.2$ & $-0.4274$ & $-0.4226$ & $1.11$ & $-0.4090$ & $-0.3722$
& $-0.3706$ & $0.41$ \\ 
$-2$ & $0.8$ & $0.3$ & $-0.4299$ & $-0.4257$ & $0.99$ & $-0.4130$ & $-0.3722$
& $-0.3709$ & $0.33$ \\ 
$-2$ & $0.8$ & $0.4$ & $-0.4320$ & $-0.4280$ & $0.93$ & $-0.4166$ & $-0.3722$
& $-0.3711$ & $0.28$ \\ 
$-2$ & $0.8$ & $0.5$ & $-0.4335$ & $-0.4297$ & $0.88$ & $-0.4197$ & $-0.3722$
& $-0.3712$ & $0.26$ \\ 
$-2$ & $0.8$ & $1.0$ & $-0.4313$ & $-0.4284$ & $0.66$ & $-0.4248$ & $-0.3722$
& $-0.3714$ & $0.20$ \\ 
$-4$ & $0.5$ & $0.1$ & $-0.6160$ & $-0.5152$ & $16.37$ & $-0.4355$ & $%
-0.5404 $ & $-0.4785$ & $11.46$ \\ 
$-4$ & $0.5$ & $0.2$ & $-0.5427$ & $-0.5067$ & $6.64$ & $-0.4412$ & $-0.5404$
& $-0.5050$ & $6.55$ \\ 
$-4$ & $0.5$ & $0.3$ & $-0.5374$ & $-0.5083$ & $5.42$ & $-0.4487$ & $-0.5404$
& $-0.5107$ & $5.51$ \\ 
$-4$ & $0.5$ & $0.4$ & $-0.5406$ & $-0.5131$ & $5.08$ & $-0.4578$ & $-0.5404$
& $-0.5129$ & $5.08$ \\ 
$-4$ & $0.5$ & $0.5$ & $-0.5440$ & $-0.5172$ & $4.92$ & $-0.4669$ & $-0.5404$
& $-0.5143$ & $4.83$ \\ 
$-4$ & $0.5$ & $1.0$ & $-0.5385$ & $-0.5168$ & $4.03$ & $-0.4897$ & $-0.5404$
& $-0.5184$ & $4.07$ \\ 
$-4$ & $0.8$ & $0.1$ & $-1.2115$ & $-1.0023$ & $17.27$ & $-0.9622$ & $%
-1.0840 $ & $-0.9534$ & $12.04$ \\ 
$-4$ & $0.8$ & $0.2$ & $-1.1042$ & $-1.0179$ & $7.82$ & $-0.9786$ & $-1.0840$
& $-1.0070$ & $7.10$ \\ 
$-4$ & $0.8$ & $0.3$ & $-1.0727$ & $-1.0162$ & $5.26$ & $-0.9924$ & $-1.0840$
& $-1.0243$ & $5.50$ \\ 
$-4$ & $0.8$ & $0.4$ & $-1.0685$ & $-1.0185$ & $4.68$ & $-1.0048$ & $-1.0840$
& $-1.0302$ & $4.96$ \\ 
$-4$ & $0.8$ & $0.5$ & $-1.0667$ & $-1.0203$ & $4.35$ & $-1.0155$ & $-1.0840$
& $-1.0338$ & $4.63$ \\ 
$-4$ & $0.8$ & $1.0$ & $-1.0462$ & $-1.0126$ & $3.21$ & $-1.0349$ & $-1.0840$
& $-1.0441$ & $3.68$%
\end{tabular}

\end{table}%
%

\section{Pseudogap formation in the density of states and in $A\left( k_{F},%
\protect\omega \right) $}

\label{section3}

In the first subsection below, we summarize previous work on pseudogap in
the attractive Hubbard model. In the second subsection, we use the approach
of section~\ref{section2a} to study the conditions under which a pseudogap
appears in the density of states and in $A(\overrightarrow{k}_{F},\omega ).$

\subsection{Overview of some recent work}

The effect of superconducting fluctuations on the density of states was
studied long ago.\cite{Abrahams:1970} To elucidate further the Physics of
pseudogap formation, especially in the single particle spectral weight, many
theoretical studies have focused on the attractive Hubbard model. An
exhaustive review may be found in Ref.\cite{Loktev:2000}. Although the
attractive Hubbard model is clearly not a realistic model for cuprates since
it predicts an $s-$wave instead of a $d-$wave superconducting ground state,
it is an extremely useful paradigm. Indeed, except for the lack of cutoff in
the interaction, it is analogous to the BCS model and is the simplest
many-body Hamiltonian that leads to superconductivity with a possible
crossover from the BCS limit at weak coupling to the Bose-Einstein limit at
strong coupling.\cite{Nozieres:1985} A key point, as far as we are
concerned, is that it also represents the only Hamiltonian for which QMC
simulations are available now as a means of checking the accuracy of
approximate many-body calculations.

The conventional picture\cite{Nozieres:1985}, based on mean-field ideas, is
that in the weak coupling regime ($\left| U/t\right| \ll 1$) pairing and
phase coherence happen at the same temperature while in the strong coupling
regime ($\left| U/t\right| \gg 1$) phase coherence may occur at $T_{c}$ much
lower than $T^{\ast }$ where pair formation happens. In the latter case, the
Fermi surface is destroyed well before the superconducting transition
occurs. Since the ARPES experiments suggest a relatively well-defined Fermi
surface, it has been suggested \cite
{Randeria:1992,Trivedi:1995,Randeria:1997} that at intermediate coupling it
is possible to retain aspects of both the Fermi surface of weak coupling and
the preformed pair ideas of strong coupling. These possibilities have been
extensively studied by several groups using a number of approaches that we
will crudely divide in two types, numerical and many-body approaches.

Previous numerical QMC work charted the phase diagram of the attractive
Hubbard model\cite{Scalettar:1989,Moreo:1991,Scalapino:1993,Assaad:1993}.
They have also investigated the pseudogap phenomenon from intermediate to
strong coupling\cite{Singer:1996,Singer:1998,Singer:1999}. We stress that
the Physics in strong coupling is different from the weak to intermediate
coupling limit we will study below. On the weak-coupling side of the BCS to
Bose-Einstein crossover, there have been numerical studies of BKT
superconductivity\cite{Moreo:1992,Assaad:1993} as well as several
discussions of pseudogap phenomena in the spin properties (susceptibility
and NMR relaxation rate) and in the total density of states at the Fermi
level.\cite{Randeria:1992,Trivedi:1995,Singer:1996} We have studied, through
Quantum Monte Carlo (QMC) simulations, the formation of a pseudogap in $%
A\left( k_{F},\omega \right) $ in the weak to intermediate coupling regime
of interest here\cite{VilkAllen:1998,Allen:1999}. The present work is in
agreement with our earlier results, as will be discussed in the next
subsection.

The many-body techniques that have been applied to the attractive Hubbard
model in the weak to intermediate coupling regime are mostly $T$-matrix and
self-consistent (Fluctuation Exchange Approximation) $T$-matrix approaches 
\cite{Serene:1989,Pedersen:1997,Schafroth:1997,Deisz:1998}. Let us consider
the pseudogap problem in the non-superconducting state. At low density\cite
{Fresard:1992,Kyung:1998}, or with additional approximations\cite
{Micnas:1995,Tchernyshyov:1997}, a pseudogap may be found. By contrast, when
the $\vec{q}=0$ superconducting mode is relaxational, self-consistent $T$%
-matrix calculations\cite{Gooding:1998} have failed to show a pseudogap in
the one-particle spectral function $A\left( k_{F},\omega \right) =-2%
\mathop{\rm Im}%
G^{R}\left( k_{F},\omega \right) $, a dimension-independent result. In
two-dimensions, this absence of pseudogap is in sharp contrast with various
QMC results\cite{VilkAllen:1998,Allen:1999} and with general physical
arguments inspired by studies of the repulsive case\cite{Vilk5.6:1997}\cite
{Preosti:1998}, which have already stressed that space dimension is crucial
in the Physics of pseudogap formation.

In the non self-consistent version of the $T$-matrix approximation however,
a pseudogap can be found\cite{Beach:1999}. Nevertheless, since the $T$%
-matrix approximation takes into account the Gaussian (first nontrivial)
fluctuations with respect to the ``mean field state'', one important
pathology of the $T$-matrix approximation in two dimensions is that the
Thouless criterion for the superconducting instability occurs at a finite
mean-field temperature $T_{MF}.$ In the very weak coupling regime, this is
considered inconsequential since the relative difference between $T_{MF}$
and the Berezinskii-Kosterlitz-Thouless\cite{Kosterlitz:1973} temperature
(BKT) $T_{BKT}$ is of order\cite{HalperinNelson:1979} $T_{MF}/E_{F}$. In the
intermediate-coupling regime however, this argument fails and one of the
questions that should be answered is precisely the size of the fluctuating
region where a pseudogap is likely to occur.

By contrast with non self-consistent $T-$matrix calculations,
self-consistent $T$-matrix approaches do have a large fluctuation region,
but they assume a Migdal theorem which means they do not take vertex
corrections into account in the self-energy formula and use self-consistent
Green functions. In the same way as in the case of repulsive interactions 
\cite{Vilk6.2:1997}\cite{Moukouri:1999}, the failure to treat vertex
corrections and fluctuations at the same level of approximation may lead to
incorrect conclusions concerning pseudogaps in $A(\overrightarrow{k}%
_{F},\omega )$. Self-consistent calculations have in fact lead to the claim 
\cite{Engelbrecht:1997} that only $d-$wave superconductivity may have
precursor effects in $A\left( k_{F},\omega \right) $ above the transition
temperature while Monte Carlo calculations\cite{VilkAllen:1998}\cite
{Allen:1999} have exhibited this pseudogap even in the $s-$wave case.
Recently, it has been pointed out using a different approach \cite
{Janko:1997} that in the intermediate-coupling regime and when the filling
is low, it becomes possible to have a bound $\vec{q}=0$ pair. This Physics
leads to a pseudogap in $A(\vec{k},\omega )$ but it requires strong
particle-hole symmetry breaking and is not specific to two dimensions. This
result is discussed further in the following subsection.

Some authors\cite{Lotkev:2000-1997,Balseiro:1992,Beck:1991-1999,Kwon:1999}
have included phenomenologically a BKT fluctuation region either in $T-$%
matrix like calculations, through Hubbard-Stratonovich transformation or
otherwise\cite{Franz:1998}. These calculations allow for phase fluctuations
in the presence of a non-zero expectation value for the magnitude of the
order parameter. In such a case, there is generally a real gap in $A(%
\overrightarrow{k}_{F},\omega ),$ and additional effects must be included to
fill-in the gap to transform it into a pseudogap\cite{Loktev:2000}. In the
approach that we take, any $SO\left( n\geq 2\right) $ theory would give
qualitatively the same result above either $T_{BKT}$ for $n=2$ or above $T=0$
for $n>2$. In addition, in our approach the magnitude of the order parameter
still fluctuates. It is when both amplitude and phase fluctuations enter the
renormalized-classical regime, {\it i.e. }become quasistatic, that a
pseudogap may open up.

\subsection{Pseudogap formation in weak to intermediate coupling}

In this section, we focus on the Physics of fluctuation-induced pseudogap in
the single-particle spectral weight of the two-dimensional attractive
Hubbard model. This Physics has been discussed in previous QMC\cite
{VilkAllen:1998,Allen:1999,Allen:2000b} and analytical work\cite
{Vilk5.6:1997} but the present quantitative approach, based on the equations
of Sec.\ref{section2a}, allows us to do calculations that are essentially in
the thermodynamic limit and that can be done sufficiently rapidly to allow
us to address other questions such as the crossover diagram in the
temperature-filling plane.

Since the cuprates are strongly anisotropic and may be considered as a quasi
2D systems, it is important to understand in detail the limiting case of
two-dimensions. Mean-field theory leads to finite-temperature phase
transitions even in low dimensional systems where breaking of continuous
symmetries is strictly forbidden at finite temperatures (Mermin-Wagner
theorem\cite{Mermin:1966}). Thus in low dimensions, mean-field theory, or
fluctuation theory based on the mean-field state, lead to qualitatively
wrong results at finite temperatures. One of the particular features of 2D
that is captured by our approach, as we will see below, is that the
mean-field transition temperature is replaced by a crossover temperature
below which the characteristic energy of fluctuations is less than
temperature, the so-called renormalized-classical regime. In this regime,
the correlation length ($\xi $) increases exponentially until, in the
superconducting case, one encounters the BKT\cite{Kosterlitz:1973}
topological phase transition. As a result we find that when $\xi \gg
v_{F}/T, $ the electronic system simulates the broken-symmetry ground state (%
$\xi =\infty $) at temperatures that are low but not necessarily very close
to the transition temperature, leading to precursors of Bogoliubov
quasiparticles\cite{Vilk5.6:1997} above $T_{c}$. Recent
dimensional-crossover studies using an analogous approach\cite{Preosti:1998}
have suggested how the pseudogap will disappear when coupling to the third
dimension is increased.

\begin{figure}%
%
\centerline{\epsfxsize 8cm \epsffile{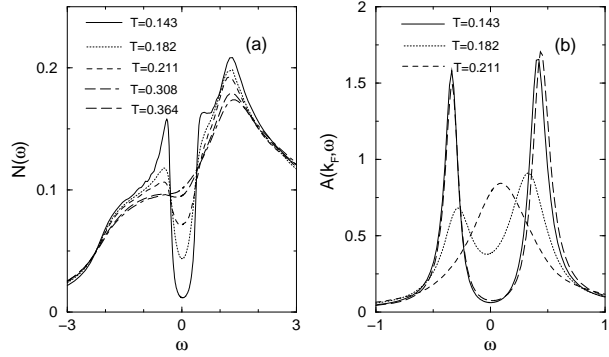}}%
%
\caption{(a) The density of states and (b) the spectral function at the
Fermi surface for $|U|=4$ and $n=0.5$ at different temperatures. The solid,
dotted, dashed, long-dashed, and dot-dashed curves correspond to $T=0.143$,
0.182, 0.211, 0.308, and 0.364, respectively, except for panel (b) 
where the long-dashed line was calculated with a $2560 \times 2560$ lattice
to illustrate that size effects are small.}%
%
\label{fig5}%
%
\end{figure}

The results of this section for two-particle properties, such as the pairing
susceptibility and the characteristic pairing fluctuation scale $\xi $, are
all computed for an effective $2560\times 2560$ lattice. The one-particle
properties are calculated instead for a $64\times 64$ lattice in momentum
space. Fast Fourier transforms (FFT) were used in that case to speed up the
calculations. The solid and long-dashed lines in Fig.\ref{fig5}(b), obtained
respectively for a $64\times 64$ and a $2560\times 2560$ lattice, illustrate
that for single-particle properties a $64\times 64$ lattice suffices. This
lattice size, for the temperatures we consider, is large enough\cite
{Moukouri:1999} compared $\xi _{th}$ (panel in Fig.\ref{fig7}(b)) that
single-particle properties are essentially the same as they would be in the
infinite-size limit. Refs.\cite{Vilk:1996} and \cite{Moukouri:1999} discuss
how single-particle properties become rather insensitive to system size even
when $\xi >L$, as long as the condition $\xi _{th}<L<\xi $ is satisfied.
This is discussed further in \cite{NoteSize}. In the calculations, equation (%
\ref{DdeRegleSomme}) is solved iteratively, then the self-energy Eq.(\ref
{Sigma(2)}) is obtained in Matsubara frequencies. The analytic continuation
from Matsubara to real frequencies are performed via Pad\'{e} approximants 
\cite{Vidberg:1977}. In order to detect any spurious features associated
with this numerical analytical-continuation, we also performed
real-frequency calculations. The results are identical, except for the fact
that the Pad\'{e} technique smooths out some of the spiky features of the
real frequency formulation that are remnants of finite-size effects when the
small imaginary part $\eta $ in retarded propagators is very small.

In Fig.~\ref{fig5} we show, for various temperatures, the total density of
states (Fig.~\ref{fig5}(a)) as well as the spectral function (Fig.~\ref{fig5}%
(b)) $A(\overrightarrow{k}_{F},\omega )$ for the Fermi surface point
crossing the $\left( 0,0\right) -\left( \pi ,0\right) $ line for $U=-4$ and
quarter filling $n=0.5$ 
%
\ For $T=0.364$ (dot-dashed curve) the density of states, on the left panel,
is similar to that for non-interacting electrons. With decreasing
temperature below $T=0.32$, the low frequency spectral weight begins to be
suppressed, leading to a pseudogap in the density of states. The condition
for the appearance of a pseudogap in the spectral function $A(%
\overrightarrow{k}_{F},\omega )$, on the right panel, is more stringent than
that in the total density of states. Although the pseudogap in the density
of states is well developed for $T=0.211$ (dashed curve), it disappears in
the spectral function for the same temperature. It is easier to form a
pseudogap in the total density of states because of its cumulative nature:
It suffices that scattering become stronger at the Fermi wave vector than at
other wave vectors to push weight away from $\omega =0.$ Hence, a pseudogap
may occur in the density of states even if $A(\overrightarrow{k}_{F},\omega
) $ remains maximum at $\omega =0.$ This is what occurs in FLEX (Fluctuation
Exchange) type calculations.\cite{Serene:1989,Deisz:1998} It is more
difficult to create a pseudogap in $A(\overrightarrow{k}_{F},\omega )$
itself since, at this wave vector, transforming a maximum at $\omega =0$ to
a minimum requires the imaginary part of the self-energy to grow very
rapidly as $T$ decreases.\cite{Vilk:1997} The generality of these arguments
suggests that $d-$wave pairing fluctuations, which were considered in Ref. 
\cite{Preosti:1998,Kyung:2000a} for example, should also lead to a pseudogap
in the density of states before a pseudogap in $A(\overrightarrow{k}%
_{F},\omega ).$ This feature is consistent with the recent experimental
observations\cite{Renner:1998} on high-temperature superconductors where
pseudogap phenomena appear at higher temperatures in tunneling experiments
than in ARPES experiments. Note also that with increasing temperature the
pseudogap in both the density of states and the spectral function appears to
fill instead of closing. This behavior is also in qualitative agreement with
tunneling \cite{Renner:1998} and with ARPES experiments\cite
{Loeser:1996,Ding:1996}. All the above results are consistent with Monte
Carlo simulations \cite{Allen:2000b}. In addition to having found a
pseudogap in the density of states \cite{Moreo:1992}, Fig.(\ref{fig2}), the
more recent Monte Carlo simulations done in the present and earlier papers 
\cite{VilkAllen:1998}\cite{Allen:1999} have also shown that a pseudogap may
occur in $A(\overrightarrow{k}_{F},\omega )$ even in $s-$wave
superconductors, contrary to the claims of Ref.\cite{Engelbrecht:1997}.

Fig. \ref{fig5}(b) also shows one other {\it qualitative} result which is a
clear signature of intermediate to strong-coupling systems, analogous to the
signatures seen in optical spectra of high-temperature superconductors\cite
{Basov:2000}. In changing $T$ by about $0.03$, from $0.21$ to $0.18,$ the
spectral weight rearranges over a frequency scale of order one, {\it i.e. }%
over a frequency scale about $30$ times larger than the temperature change,
and $5$ times larger than the absolute temperature $0.2$. In weak-coupling
BCS theory, by contrast, spectral weight rearranges over a frequency scale
of the order of the temperature change. The frequency range for spectral
rearrangement observed in Fig. \ref{fig5}(b) would be even larger if the
coupling was stronger.\cite{Pairault:2000} This is a consequence of the fact
that wave vector can be a very bad quantum number for correlated systems so
that a momentum eigenstate can project on essentially all the true
eigenstates of the system. The loss of meaning of momentum as a good quantum
number and the corresponding spectral weight rearrangement over a large
frequency scale happens suddenly in temperature in Fig. \ref{fig5}(b)
because the correlation length becomes large at a rather sharp threshold
temperature where the system becomes renormalized classical, as we now
discuss.

\begin{figure}%
%
\centerline{\epsfxsize 8cm \epsffile{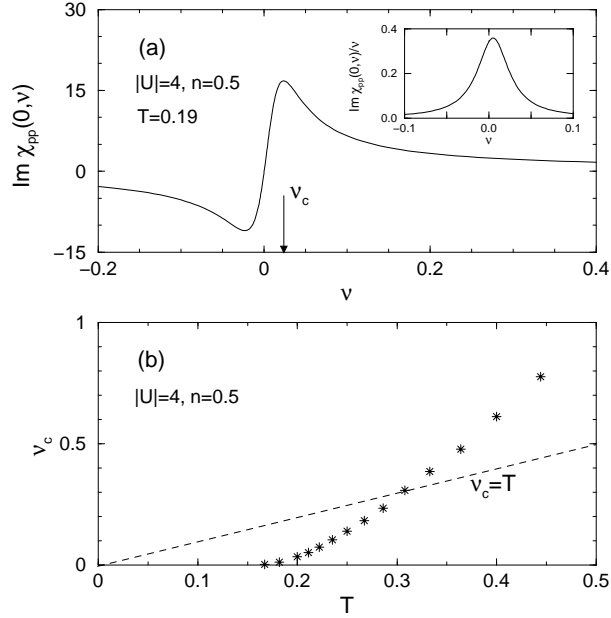}}%
%
\caption{(a) The imaginary part of the pairing susceptibility at $\vec{q}=0$
for $T=0.19$ and (b) the characteristic low energy scale (stars) for pairing
fluctuations at different temperatures for $|U|=4$ and $n=0.5$. The inset in
(a) is the imaginary part of the pairing susceptibility divided by frequency
at $\vec{q}=0$ for $T=0.19$.}%
%
\label{fig6}%
%
\end{figure}

\begin{figure}%
%
\centerline{\epsfxsize 8cm \epsffile{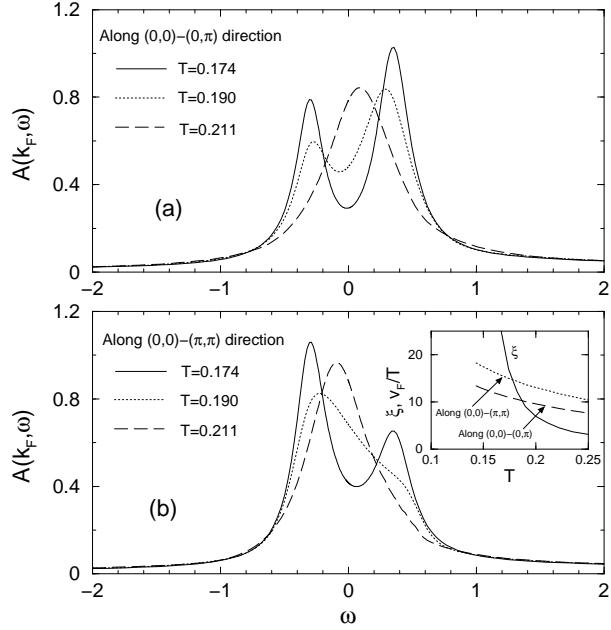}}%
%
\caption{Spectral function for $|U|=4$ and $n=0.5$ (a) along $(0,0)-(0,\pi )$ 
direction and (b) along $(0,0)-(\pi ,\pi )$
direction. The inset in (b) shows the pairing correlation length (solid
curve), and $\xi_{th}=v_{F}/T$ along $(0,0)-(\pi ,\pi )$ direction
(dotted curve) and $(0,0)-(0,\pi )$ direction (dashed curve) for the
same parameters.}%
%
\label{fig7}%
%
\end{figure}

Let us then demonstrate that the opening of the pseudogap in the
single-particle density of states occurs when the pairing fluctuations enter
the renormalized-classical regime.\cite{VilkAllen:1998} In Fig.~\ref{fig6}%
(a) the imaginary part of the paring susceptibility at $\vec{q}=0$ for $%
T=0.19$ and the characteristic frequency $\nu _{c}$ for pairing fluctuations
are shown for $U=-4$ and $n=0.5$.%
%
\ Since for the parameters studied here the $\vec{q}=0$ mode is deep in the
particle-particle scattering continuum, it has the characteristic frequency
dependence of a relaxational mode, $1/(1-i\nu /\nu _{c}),$ that leads to a
maximum in the imaginary part at some characteristic frequency $\nu _{c}$.
Even though we do not have perfect particle-hole symmetry, the Fermi energy
is still large enough compared with temperature that $%
\mathop{\rm Im}%
\chi _{pp}\left( 0,\nu \right) /\nu $ is very nearly even (Inset in Fig.~\ref
{fig6}(a)). For other temperatures the behavior is similar. In Fig.~\ref
{fig6}(b) $\nu _{c}$ is plotted as a function of temperature. At high
temperatures $\nu _{c}$ is larger than $T$ but below $T\simeq 0.31-0.32$ the
characteristic frequency $\nu _{c}$ becomes smaller than $T$, signaling that
we are entering the renormalized-classical regime. This phenomenon was also
observed in QMC calculations.\cite{VilkAllen:1998} In this regime, the
thermal occupation number for pairing fluctuations is larger than unity.
Clearly, the appearance of a pseudogap in the density of states in Fig.~\ref
{fig5}(a) follows very closely the entrance in the renormalized classical
regime.

The present results should be contrasted with those of Levin's group\cite
{Janko:1997}. The pseudogap in their work comes from the presence of a $\vec{%
q}=0$ resonant pair state in the $T$-matrix. As the interaction strength
decreases or the particle density increases, the $\vec{q}=0$ bound state
enters into the particle-particle continuum, thereby acquiring a finite
lifetime. As long as the $\vec{q}=0$ pair state is near the bottom of the
scattering continuum it can remain a resonant state with a relatively long
lifetime. Thus the origin of a pseudogap in their study is analogous to the
preformed-pair scenario where the $\vec{q}=0$ pair is separated from the
scattering continuum. Such a resonance corresponds to strong particle-hole
asymmetry in the imaginary part of the pair susceptibility. In order to have
such an asymmetry for moderate coupling strength, very small particle
density is required in this approach. In our case, the pseudogap occurs even
when the particle-hole symmetry is nearly perfect. Furthermore, in our case,
other factors like density and interaction strength do not influence the
results in any dramatic way. Low dimensionality is the key factor since
phase space is behind the existence of both the renormalized classical
regime and the very strong scattering of electrons on the corresponding
fluctuations. The ratio $\xi /\xi _{th}$ controls the importance of this
scattering\cite{Vilk:1996} as we discuss in the following paragraph.

In Fig.~\ref{fig7} we contrast the onset of the pseudogap in the spectral
function on the Fermi surface along different directions, namely the $%
(0,0)-(0,\pi )$ and $(0,0)-(\pi ,\pi )$ directions, for $|U|=4$ and $n=0.5$. 
%
\ At this density, where the Fermi surface is nearly circular, the
anisotropy happens in a very small temperature range around $T=0.19$. For $%
T=0.19$ (dotted curves), the figure shows that the pseudogap occurs only
along the $(0,0)-(0,\pi )$ direction. This anisotropy of the pseudogap in
the spectral function should be contrasted with the fact that in the
superconducting state, the gap is isotropic. The anisotropy at the
temperature where the pseudogap opens up can be understood following the
arguments of Vilk {\it et al.}\cite{Vilk:1997}. Using the dominant
renormalized-classical fluctuations ($iq_{n}=0$), these authors showed that
for $\xi \gg v_{F}/T$ the scattering rate (imaginary part of the
self-energy) on the Fermi surface becomes large, leading to a minimum in the
spectral function at $\omega =0$ instead of the maximum that exists in the
absence of a pseudogap. In the inset of Fig.\ref{fig7}(b) the pairing
correlation length $\xi $, as well as $\xi _{th}=v_{F}/T$ along the $%
(0,0)-(\pi ,\pi )$ and $(0,0)-(0,\pi )$ directions are plotted as a function
of temperature. Clearly $\xi $ grows exponentially with decreasing
temperature. Furthermore, according to the above criterion, a pseudogap in
the spectral function exists along one direction and not along the other
when $\xi $ (solid curve) is larger than $\xi _{th}$ along $(0,0)-(0,\pi )$
(dashed curve) but smaller than $\xi _{th}$ along the $(0,0)-(\pi ,\pi )$
(dotted curve), namely, in the temperature range $0.175<T<0.185$. We obtain
quantitative agreement with Fig.~\ref{fig7}(b) if we use $\xi =1.3\xi _{th}$
as the criterion for the appearance of a pseudogap. While the pseudogap
anisotropy happens in a narrow temperature range at this density due to the
small Fermi velocity anisotropy (about $1.35$), closer to half-filling it
occurs in a large temperature interval since the Fermi velocity is nearly
vanishing close to the $(0,\pi )$ point.\cite{Allen:1999}

\begin{figure}%
%
\centerline{\epsfxsize 8cm \epsffile{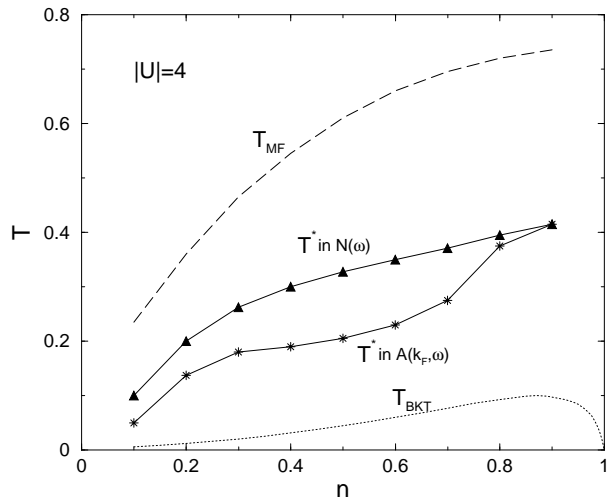}}%
%
\caption{The crossover diagram of the 2D attractive Hubbard model for $|U|=4$.
The filled triangles and stars denote the temperatures where a pseudogap
appears in the density of states and the spectral function, respectively.
The solid lines are a guide to the eye. The dashed curve is the BCS mean field
temperature $T_{MF}$ and the dotted curve is an estimate of the Kosterlitz-Thouless
temperature $T_{KT}$ extracted from QMC results by Moreo {\it et al.}
\protect\cite{Moreo:1991}.}%
%
\label{fig8}%
%
\end{figure}

Finally in Fig.~\ref{fig8} we present the crossover diagram for the
pseudogap in the 2D attractive Hubbard model for $|U|=4$.%
%
\ The dotted curve is a rough QMC\cite{Moreo:1991} estimate (probably an
upper bound) for the BKT transition temperature $T_{BKT}$. For all densities
a pseudogap in the one-particle functions appears in a wide temperature
range $T_{BKT}<T<T^{\ast }$ where $T^{\ast }$ is typically several times of $%
T_{BKT}=T_{c}$. The pseudogap occurs earlier in the density of states than
in the spectral functions for most of densities. Near half-filling, however,
the pseudogap appears more or less at the same temperature in the density of
states and the spectral functions. In QMC for small systems, there seems to
be a difference in the temperatures at which the two pseudogaps open up.\cite
{Allen:2000b} Performing a calculation with finite second neighbor hopping $%
t^{\prime },$ we have confirmed that this almost simultaneous opening of the
pseudogaps happens because of the strong influence of the Van Hove
singularity, which leads to $v_{F}=0,$ and not because of nesting.\cite
{Notet'} Finally, note that at half-filling one has perfect $O\left(
3\right) $ symmetry in this model so that the transition temperature
vanishes, as dictated by the Mermin-Wagner theorem, while the pseudogap
temperature continues to be large, following the trend of the mean-field
transition temperature instead of that of the $T_{BKT}$ curve. This shows
that symmetry of the order-parameter space contributes to enlarge the
temperature range where the pseudogap occurs, as expected from the
corresponding enlargement of the renormalized-classical regime.\cite
{Allen:1999}

\section{Conclusion}

\label{section4}

In weak to intermediate coupling, the attractive Hubbard model can be
studied quantitatively with a non-perturbative approach\cite{Allen:2000}
that directly extends the corresponding method for the repulsive model\cite
{Vilk:1994,Vilk:1996,Vilk:1997}. The simple equations of section \ref
{section2a} are all that needs to be solved. This many-body approach has an
internal accuracy check, no adjustable parameter and it satisfies several
exact sum rules\cite{Allen:2000}. We have demonstrated the accuracy of this
method through detailed comparisons of its predictions with Quantum Monte
Carlo simulations of both single-particle and two-particle correlation
functions.

On the Physical side, we studied the fluctuation-induced pseudogap that
appears in the single-particle spectral weight, in agreement with Monte
Carlo simulations and in close analogy with the results found before in the
repulsive case\cite{Vilk:1997}\cite{Moukouri:1999}. A key ingredient for
this pseudogap is the low dimensionality. Indeed, in two-dimensions the
finite-temperature mean-field transition temperature is replaced by a
crossover to a renormalized-classical regime where the characteristic
pairing frequency is smaller than temperature and the pairing correlation
length $\xi $ grows faster than the single-particle thermal de Broglie
wavelength $\xi _{th}$. In this approach, where vertex corrections and Green
functions are taken at the same level of approximation in the self-energy
expression\cite{Vilk:1997}\cite{Moukouri:1999}, the renormalized-classical
fluctuations and the relatively large phase space available for them in
two-dimensions lead to precursors of the superconducting gap (or Bogoliubov
quasiparticles) in the normal state. This pseudogap can occur without
resonance in the pair susceptibility\cite{Janko:1997}, and it appears not
only in the total density of states but also in the single-particle spectral
weight, in sharp contrast with what was found with self-consistent $T-$%
matrix approaches\cite{Gooding:1998}. Our approach fails at strong coupling
or at low temperature very close to the Berezinskii-Kosterlitz-Thouless
(BKT) transition.

For $\left| U\right| =4,$ the pseudogap regime occurs over a temperature
scale that is several times the BKT transition temperature. The crossover to
the renormalized-classical regime is about a factor of two lower than the
mean-field transition temperature but it has the same filling dependence,
which can be quite different from that of the real transition temperature,
which is strongly dependent on the symmetry of the order-parameter space\cite
{Allen:1999}. It is clear also that $SO\left( 2\right) $ (or $U\left(
1\right) $) symmetry is not essential to the appearance of a pseudogap. It
would also appear if there happens to be a hidden continuous symmetry group 
\cite{Zhang:1997}\cite{Allen:1999} $SO\left( n\right) $ with $n\geq 2$
describing the high-temperature superconductors.

As stressed earlier in this paper, the attractive Hubbard model is not
directly applicable to the cuprates. Nevertheless, it helps understand the
nature of superconducting-fluctuation induced pseudogaps, if they happen to
be present. The pseudogap appearing for the {\it underdoped} compounds at 
{\it high} temperature in thermodynamic and transport measurements, or at
high energy in tunnelling\cite{Renner:1998} and ARPES experiments, is most
probably {\it not} of {\it pure} superconducting origin\cite{Loram:2000}\cite
{Krasnov:2000}. Nevertheless, close enough to the superconducting
transition, in {\it both} the {\it underdoped} and {\it overdoped} regions,
there should be an effective model with attraction describing the low energy
Physics. Since even the high-temperature superconductors have a gap to Fermi
energy ratio that is small, this effective model could be a weak-coupling
one (but not necessarily\cite{Note:FaibleVsFort}). Time-domain transmission
spectroscopy experiments\cite{Orenstein:1998} in the $100GHz$ range suggest
that the renormalized classical regime for the BKT transition has been
observed in {\it underdoped} compounds, $10K$ to $15K$ above $T_{c}$. Also,
in the {\it overdoped} regime, recent experiments on the magnetic field
dependence of NMR $T_{1}^{-1}$ and Knight shift\cite{Zheng:2000} suggest
that the pseudogap appearing a few tens of degrees above $T_{c}$ is indeed a
superconducting-fluctuation induced pseudogap. The pseudogap that we have
described should appear in these regimes if an effective weak to
intermediate-coupling attractive-interaction model is valid near $T_{c}$. In
this context, some of the important results that we found and explained are
as follows. In the attractive Hubbard model the pseudogap appears earlier in
the density of states than in the spectral function that would be measured
by ARPES, as summarized in Fig.(\ref{fig8}). We also found, Fig.(\ref{fig5}%
), that with increasing temperature, spectral weight appears to fill in the
pseudogap instead of closing it. Finally, we also showed that as the system
enters the renormalized-classical regime, spectral weight can rearrange over
a frequency range much larger than the temperature scale. This is generally
a signature that momentum is becoming a very bad quantum number. Hence, for
a given temperature scale, the frequency range over which the spectral
weight can rearrange becomes larger with increasing coupling\cite
{Pairault:2000}. All these features carry over the $d-$wave case\cite
{Preosti:1998,Kyung:2000a}. Qualitative differences between weak- and
strong-coupling pseudogaps have been discussed in Ref.\cite{Moukouri:1999}.

\acknowledgements

A.-M.S.T. is indebted to Y.M. Vilk for numerous important discussions and
suggestions. We also thank D. Poulin and S. Moukouri for their Maximum
Entropy code, H. Touchette for invaluable help with the QMC code and F.
Lemay for numerous discussions and for sharing the results of some of his
calculations. The authors would like to thank R. Gooding, F. Marsiglio and
M. Capezzali for useful discussions. A.M.S.T. would also like to thank E.
Bickers and P. Hirschfeld for discussions. Monte Carlo simulations were
performed in part on IBM-SP computers at the Centre d'Applications du Calcul
Parall\`{e}le de l'Universit\'{e} de Sherbrooke. This work was partially
supported by the Natural Sciences and Engineering Research Council of Canada
(NSERC), by the Fonds pour la Formation de Chercheurs et l'Aide \`{a} la
Recherche (FCAR) from the Qu\'{e}bec government, the Canadian Institute for
Advanced Research and in part, at the Institute for Theoretical Physics,
Santa Barbara, by the National Science Foundation under grand No.
PHY94-07194.


\begin{references}
\bibitem{Note:TPSC}  The name Two-Particle Self-Consistent (TPSC) is because
the self-consistency in Eq.(\ref{DdeRegleSomme}) involves the
particle-particle irreducible vertex and a special case of the pair
susceptibility. Consistency with $\Sigma ^{\left( 1\right) }$ for this
Hamiltonian is trivial but it would also need to be taken into account for
models more general than Hubbard's.
\bibitem{Note:FaibleVsFort}  Even though the attractive part of the
Hamiltonian could be in the weak coupling regime, if there is a strongly
repulsive term also in the Hamiltonian the details of the Physics near the
superconducting transition could be quite different from what is found in
the attractive Hubbard model. Nevertheless, the fact that we were able to
treat the model up to intermediate coupling suggests that many of our
qualitative results should hold in the more general case.
\end{references}
\end{document}